# *M|D|∞* Queue Busy Period and Busy Cycle Distributions Computational Calculus

**Manuel Alberto M. Ferreira**

manuel.ferreira@iscte-iul.pt

**Abstract**

Given the busy period and busy cycle major importance in queuing systems, it is crucial the knowledge of the respective distribution functions that is what allows the calculation of the important probabilities. For the $M|G|\infty$ queue system, there are no round form formulae for those distribution functions. But, for the $M|D|\infty$ queue, due the fact that its busy period and busy cycle have both Laplace transform expression round forms, what does not happen for any other $M|G|\infty$ queue system, with an algorithm created by Platzman, Ammons and Bartholdi III, that allows the tail probabilities computation since the correspondent Laplace transform in round form is known, those distribution functions calculations are possible. Here, we will implement the algorithm through a FORTRAN program.

**1 Introduction**
A queue system busy period is a period that begins when a customer arrives at the system finding it empty and ends when a customer abandons the system letting it empty. Throughout its progress, there is always at least one customer present. In any queue operation there is an alternate sequence of idle and busy periods. An idle period followed by a busy period is a busy cycle.

In the $M|G|\infty$ queue system the customers arrive according to a Poisson process at rate $\lambda$, receive a service which time length is a positive random variable with distribution function $G(.)$ and mean $\alpha$. When they arrive, each one finds immediately an available server. Each customer service is independent from the other customers' services and from the arrivals process. The traffic intensity is

$$\rho = \lambda\alpha \qquad (1.1)$$

Call $B$ the busy period time length random variable, $b(t)$ the correspondent probability density function, and $B(t)$ the distribution function.

Being $\bar{B}(s)$ the $B$ Laplace transform

$$\bar{B}(s) = 1 + \lambda^{-1}\left(s - \frac{1}{\int_0^\infty e^{-st-\lambda\int_0^t[1-G(v)]dv}dt}\right) \quad (1.2),$$

see [2].

Consequently, see [6],

$$E[B^n] = (-1)^{n+1}\left\{\frac{e^\rho}{\lambda} n\, C^{(n-1)}(0) - e^\rho \sum_{p=1}^{n-1}(-1)^{n-p}\binom{n}{p}E[B^{n-p}]C^{(p)}(0)\right\},$$

$$n = 1,2,\ldots \quad (1.3)$$

and

$$C^{(n)}(0) = \int_0^\infty (-t)^n e^{-\lambda\int_0^t[1-G(v)]dv}\lambda(1-G(t))dt, n = 0,1,2,\ldots \quad (1.4).$$

So,

$$E[B] = \frac{e^\rho - 1}{\lambda} \quad (1.5)$$

does not depend on the service time distribution form, except for its mean[1]. And

For the $M/D/\infty$ queue system – constant[2] service times with value $\alpha$ –

$$\bar{B}(s) = 1 + \lambda^{-1}\left(s - \frac{(s+\lambda)s}{\lambda e^{-(s+\lambda)\alpha} + s}\right) \quad (1.6)$$

---

[1] In these circumstances it is usual to say that it is insensible to the service time distribution.

[2] That is: **D**eterministic service times.

obtaining, by Laplace transform inversion, see [3][3],

$$b(t) = \sum_{n=0}^{\infty} \left(\frac{d}{dt}\frac{c(t)}{e^{-\rho}}\right) * \left(\frac{d}{dt}\frac{1-d(t)}{1-e^{-\rho}}\right)^{*n} e^{-\rho}(1-e^{-\rho})^n \quad (1.7)$$

where $\frac{c(t)}{e^{-\rho}} = \begin{cases} 0, t < \alpha \\ 1, t \geq \alpha \end{cases} = G(t)$ and $\frac{1-d(t)}{1-e^{-\rho}} = \begin{cases} \frac{1-e^{-\lambda t}}{1-e^{-\rho}}, t < \alpha \\ 1, t \geq \alpha \end{cases}$. And

$$VAR[B] = \frac{e^{2\rho} - 2\rho e^{\rho} - 1}{\lambda^2} \quad (1.8).$$

The expression (1.7) for $b(t)$, allows the busy period distribution structure interpretation for the $M|D|\infty$ queue. But it fails in the task of presenting an easy expression for the distribution function $B(t)$ computation. This may be done, for example, with an algorithm created by Platzman, Ammons and Bartholdi III, see [1], that allows the tail probabilities computation since the correspondent Laplace transform in round form is known, as it is now the case, remember (1.6), that will be the subject of next section. Unhappily the same does not happen for other $M|G|\infty$ systems what inhibits the use of this algorithm. In section 3 we will present practical applications of this case.

The same problem occurs with the calculation of the busy cycle distribution function, and the procedure described above is a way to solve it since the Laplace transform of the busy cycle also has a round form.

So, call $I$, and $Z$ the time length random variable of the idle period, and the busy cycle respectively; $I(t)$, and $Z(t)$ the distribution functions. Evidently, $Z = I + B$ and being I and B independent, see [2], the distribution of $Z$ is the $I$ and $B$ distributions convolution. Then, being $\bar{Z}(s)$, and $\bar{I}(s)$ the Z, and I, respectively, Laplace transforms:

---

[3] * is the convolution operator.

$$\bar{Z}(s) = \bar{I}(s)\,\bar{B}(s) \quad (1.9)$$

where

$$\bar{I}(s) = \frac{\lambda}{\lambda + s} \quad (1.10)$$

as it happens for any queue with Poisson (note the *M* in *M*/*D*/∞ arrivals process and consequently:

$$E[Z^n] = \sum_{p=0}^{\infty} \binom{n}{p} \frac{p!}{\lambda^p} E[B^{n-p}], n = 1,2,\ldots \quad (1.11).$$

So

$$E[Z] = \frac{e^\rho}{\lambda} \quad (1.12).$$

does not depend on the service time distribution form, except for its mean. But $E[Z^n], n \geq 2$ depend on the whole service time distribution structure. So, for the *M*/*D*/∞ queue system:

$$VAR[Z] = \frac{e^{2\rho} - 2\rho e^\rho}{\lambda} \quad (1.13).$$

This subject will be addressed in section 4. We will close this work with a brief conclusions section.

## 2 Algorithm Implementation to Compute the *M*|*D*|∞ Queue Busy Period Distribution Function

It is generally said that an algorithm is "accurate" if it looks for solving a problem "close" to the one that is supposed to solve. An algorithm is "precise" if it gets a solution "close" to the one of the problem that it is trying to solve. More concretely, being $\Delta t$ ($\Delta t > 0$) the accuracy and $\Delta p$ $\left(0 < \Delta p < \frac{1}{2}\right)$ the precision required, the approximation $\tau$ of $P[X > t]$ must satisfy the condition:

$$P[X \geq t + \Delta t] - \Delta p \leq \tau \leq P[X > t - \Delta t] + \Delta p \quad (2.1).$$

Platzman, Ammons and Bartholdi III, see [1], suggest doing

$$\tau = \frac{U - t + \Delta t}{U - L + 2\Delta t} + \sum_{n=1}^{N} \frac{\alpha^{n^2}}{\pi n} im\{(\beta^n - \gamma^n) L(j\omega n)\} \quad (2.2)$$

where $K = \log \frac{2}{\Delta p}$, $D = \frac{\Delta t}{\sqrt{2K}}$, $\omega = \frac{2\pi}{U-L+2\Delta t}$, $N = \left[\frac{2K}{\omega \Delta t}\right]$, being $[\cdot]$ the characteristic of a real number, $\alpha = e^{-D^2 \frac{\omega^2}{2}}$, $\beta = e^{j(U+\Delta t)\omega}$, $\gamma = e^{jt\omega}$, $U$ and $L$ are numbers such that $1 - P[L \leq X \leq U] \ll \Delta p$, $j = \sqrt{-1}$ and $im(\cdot)$ designates the imaginary part of a complex number. $L(j\omega n)$ is the Laplace transform value in $j\omega n$. They demonstrate that the approximation so defined fulfills the condition (2.1).

In general terms this method can be described as follows:
- To ensure fast execution, only $N$ values of the transform are calculated. These values are carefully selected to ensure as much information as possible. The exact value of the tail corresponding to the smoothest distribution function which transforms passes through these $N$ points is then calculated.
- Such a method is expected to behave at least as well as any other method that calculates $N$ values of the transform, and any other algorithm that has you calculate it more times.
- In [1], the authors also show that calculating a tail from a transform is a problem with difficulty level #P-hard. This is indicative of the computational effort required because solving a #P-hard problem, even with only a certain guarantee of approximation, requires an additional calculation that grows exponentially with the description of the problem. Note that the algorithm provides a solution not to the original problem but to an approximation defined by $\Delta t$ and $\Delta p$.
- Note that in the error definition used, $\Delta t$ refers to a perturbation of the parameter $t$ while the more common definition of error refers to a $\Delta p$ perturbation of the result.

We can apply this algorithm to calculate the distribution functions of the $M|D|\infty$ queue busy period and the busy cycle because, in these cases, both Laplace transforms have simple forms. Let's look at the case of the busy period:

$$\bar{B}(s) = 1 + \lambda^{-1}\left(s - \frac{(s+\lambda)s}{\lambda e^{-(s+\lambda)a} + s}\right) \qquad (2.3)$$

as in (1.6) but with $a$ for the service value instead of $\alpha$ for obvious reasons.

Let's start by noting that through Chebyshev's inequality:

$$P(|X - \mu| \geq K\sigma) \leq \frac{1}{K^2} \qquad (2.4),$$

being $X$ a random variable such that $E[X] = \mu$ and $VAR[X] = \sigma^2$. But $P(|X - \mu| \geq K\sigma) \leq \frac{1}{K^2} \Leftrightarrow P(X - \mu \leq -K\sigma \vee X - \mu \geq K\sigma) \leq \frac{1}{K^2} \Leftrightarrow P(X \leq \mu - K\sigma) + P(X \geq \mu + K\sigma) \leq \frac{1}{K^2}$. So, supposing that

- $X$ is a positive random variable,
- $\mu - K\sigma < 0$,
- $\mu + K\sigma = t$,

as $K = \frac{t-\mu}{\sigma}$ and so $\mu - K\sigma < 0 \Leftrightarrow \mu - \frac{t-\mu}{\sigma}\sigma < 0 \Leftrightarrow t > 2\mu$,

$$P(X \geq t) \leq \frac{\sigma^2}{(X - \mu)^2}, \text{ since } t \geq 2\mu \quad (2.5).$$

The bound given in expression (3.5) will be of interest since $\frac{\sigma^2}{(t-\mu)^2} < 1 \Leftrightarrow t < \mu - \sigma \vee t > \mu + \sigma$.

For the $M/D/\infty$ queue busy period, it will be:

$$\mu = \frac{e^\rho - 1}{\lambda}$$
$$\sigma^2 = \frac{e^{2\rho} - 2\rho e^\rho - 1}{\lambda^2} \quad (2.6),$$

confer with (1.5) and (1.8).

And being $B^D(t)$ its distribution function:

If $t > \lambda^{-1}\left[e^\rho - 1 + \max\left[e^\rho - 1; \sqrt{e^{2\rho} - 2\rho e^\rho - 1}\right]\right], B^D(t) \geq B_1^D(t)$, being $B_1^D(t) = 1 - \frac{e^{2\rho} - 2\rho e^\rho - 1}{(1+\lambda t - e^\rho)^2}$ (2.7)

So, to apply the algorithm to calculate the $M/D/\infty$ queue's busy period distribution function, we will have:

- $L = a$,
- $U = \lambda^{-1}\left(e^\rho - 1 + \sqrt{\frac{e^{2\rho} - 2\rho e^\rho - 1}{\Delta p} 10^l}\right)$, l=1,2, ...,

because

$$1 - P(L \leq X \leq U) = 1 - P(a \leq X \leq U) = 1 - P(0 \leq X \leq U),$$

having to be $B^D(U) > 1 - 10^{-l}\Delta p$, this happening if

$$\frac{e^{2\rho} - 2\rho e^\rho - 1}{(1+\lambda U - e^\rho)^2} = 10^{-l}\Delta p \Leftrightarrow (1 + \lambda U - e^\rho)^2 = \frac{e^{2\rho} - 2\rho e^\rho - 1}{\Delta p} 10^l,$$

which leads to the indicated result,

- $B^D(t) \geq B_2^D(t) = \begin{cases} 0, t < \alpha \\ e^{-\rho}, t \geq \alpha \end{cases}$,

- $t$(time),

- The desired values $B_C^D(t)$, are given by $1 - \tau$.

Making $l = 3$, we build the computer program in FORTRAN language to implement the algorithm (it is necessary to indicate the values of *a*, *t*, $\Delta t$, and $\Delta p$) that follows:

```
PROGRAM TPROG

REAL      T, DELTA, DELTP, APEQ, LAMBDA, RO
REAL      KAPA, D, OMEGA, ALFA, U, PI, TAU, X, Y, SOMA, XX, XXX
COMPLEX   BETA, GAMA, CC, CLAMBD, CAPEQ, CL
INTEGER   N, I

DATA      PI/3.14157/

PRINT *, 'T '
READ *, 'T
PRINT *, 'APEQ '
READ *, 'APEQ
PRINT *, 'LAMBDA '
READ *, 'LAMBDA
PRINT *, 'DELTA '
READ *, 'DELTA '
PRINT *, 'DELTP '
READ *, DELTP

RO =   LAMBDA* APEQ
U   =  EXP(2*RO) - 2*RO*EXP(RO)-1
U   = (U/ DELTP) *1000)
U   =  SQRT(U)
U   =  EXP (RO) -1 + U
U   =  U/LAMBDA

PRINT *, 'U
PAUSE
```

```
KAPA  = LOG(2/DELTP)
D     = DELTA/SQRT(2*K)
OMEGA = 2*PI / (U-APEQ+2*DELTA)

N = NINT (2*KAPA/(DELTA*OMEGA))

PRINT *,'N = ', N
PAUSE

ALFA = EXP (-(D*OMEGA) **2/2)

X = COS ((U+DELTA) *OMEGA)
Y = SIN ((U+DELTA) *OMEGA)
BETA = CMPLX (X, Y)

X = COS (A*OMEGA)
Y = SIN (A*OMEGA)
GAMA = CMPLX (X, Y)

SOMA = 0
DO 100 I=1, N
      X      = OMEGA * REAL (I)
      CC     = CMPLX (0.0, X)
      CLAMBD = CMPLX (LAMBDA, 0.0)
      CAPEQ  = CMPLX (APEQ, 0. 0)
      CL     = CC* (CC + CLAMBD)
      CL = CL/ (CLAMBD* CEXP (-(CC+CLAMBD) * CAPEQ) +CC)
      CL= (CC-CL) – CLAMBD + CMPLX (1,0)
      CL= (BETA**I-GAMA**I) * CL

      X = AIMAG (CL)

      XXX= PI * REAL (I)

      Y= ((ALFA ** REAL (I)**REAL (I)) /(XXX)

      SOMA=SOMA + Y*X

100   CONTINUE

      TAU= 1- (U-A+DELTA) /(U-APEQ+2*DELTA) -SOMA
```

PRINT *, 'TAU=', TAU

        STOP

        END

## 3 $M|D|\infty$ Queue Busy Period Distribution Function Computation

In this section, we present the results of applying the algorithm to calculate the distribution function of the busy period of the $M|D|\infty$ system, in the following cases:

- 1. $\alpha = .1$ and $\lambda = 1$ (Table 3.1)
- 2. $\alpha = 1$ and $\lambda = 1$ (Table 3.2)
- 3. $\alpha = 1$ and $\lambda = 1$ (Table 3.3)

We compare the values of $B_C^D(t)$ obtained with those of the lower boundaries: $B_1^D(t) = 1 - \frac{e^{2\rho} - 2\rho e^{\rho} - 1}{(1+\lambda t - e^{\rho})^2}$ and $B_2^D(t) = \begin{cases} 0, t < \alpha \\ e^{-\rho}, t \geq \alpha \end{cases}$ (see former section).

**Table 3.1**

$\alpha = .1 \ \lambda = 1 \ \rho = .1 \ \Delta t = .001 \ \Delta p = .001$

| $t$ | $B_1^D(t)$ | $B_2^D(t)$ | $B_C^D(t)$ |
|---|---|---|---|
| .1 | -12.784463 | .904837 | .453519 |
| .11 | -14.805955 | .904837 | .91431 |
| .15 | .316597 | .904837 | .950782 |
| .2 | .959013 | .904837 | .996209 |
| .25 | .982428 | .904837 | .999575 |
| | | | |
| CALCULTIONS | EXACT | Calculated from $B_C^D(t)$ with $B_C^D(.1) = .904837$ | ERROR |
| $E[B]$ | .105170918 | .1049714128 | .2% |
| $VAR[B]$ | .0003685744 | .00031661238 | 14% |

**Table 3.2**

$$\alpha = 1 \ \lambda = 1 \ \rho = 1 \ \Delta t = .1 \ \Delta p = .001$$

| $t$ | $B_1^D(t)$ | $B_2^D(t)$ | $B_C^D(t)$ |
|---|---|---|---|
| 1 | -21.921031 | .367879 | .190999 |
| 2 | -148.002717 | .367879 | .741497 |
| 3 | -6.198447 | .367879 | .907228 |
| 4 | -1.271433 | .367879 | .969885 |
| 5 | -.098048 | .367879 | .992784 |
| | | | |
| CALCULTIONS | EXACT | Calculated from $B_C^D(t)$ with $B_C^D(1) = .367879$ | ERROR |
| $E[B]$ | 1.718281828 | 1.6649785 | 3% |
| $VAR[B]$ | .9524924414 | .70343785 | 26% |

**Table 3.3**

$$\alpha = 3 \ \lambda = 1 \ \rho = 3 \ \Delta t = .5 \ \Delta p = .01$$

| $t$ | $B_1^D(t)$ | $B_2^D(t)$ | $B_C^D(t)$ |
|---|---|---|---|
| 3 | -.0895519 | .0497871 | .025126 |
| 4 | -.238790 | .0497871 | .099527 |
| 5 | -.420929 | .0497871 | .148885 |
| 6 | -.646402 | .0497871 | .198405 |
| 7 | .930133 | .0497871 | .244893 |
| 8 | -1.294064 | .0497871 | .288204 |
| 9 | -1.771539 | .0497871 | .329391 |
| 10 | -2.415214 | .0497871 | .368208 |
| 15 | -15.889655 | .0497871 | .530699 |
| 20 | -336.121704 | .0497871 | .65134 |
| 25 | -7.0691347 | .0497871 | .740937 |
| 30 | -1.366543 | .0497871 | .807469 |
| 35 | -.113102 | .0497871 | .856896 |
| 40 | .355496 | .0497871 | .893608 |
| 45 | .580208 | .0497871 | .920880 |
| 50 | .705018 | .0497871 | .941125 |
| 55 | .781435 | .0497871 | .956144 |
| 60 | .831591 | .0497871 | .967298 |
| 70 | .891248 | .0497871 | .981726 |
| 75 | .909828 | .0497871 | .986298 |
| 80 | .924024 | .0497871 | .989706 |
| 85 | .935113 | .0497871 | .992233 |
| | | | |
| CALCULTIONS | EXACT | Calculated from $B_C^D(t)$ with $B_C^D(3) = .0497871$ | ERROR |
| $E[B]$ | 19.08553692 | 18.60845683 | 2% |
| $VAR[B]$ | 281.9155718 | 250,9405890 | 11% |

The values of $B_C^D(t)$ always satisfy those of $B_1^D(t)$, which are sometimes trivial, and those of $B_2^D(t)$ except only for $t=.1$, $t=1$, and $t=3$ in Tables 3.1, 3.2, and 3.3 respectively.

Note that the busy period of this queue system has a probability concentration at $t = \alpha$ of $e^{-\rho}$ value[4]. Thus, to test the validity of the values obtained, we calculated the mean and variance from the $B_C^D(t)$ values obtained, but considering $B_C^D(\alpha) = e^{-\rho}$, and compared their values with the true ones.

The values obtained for the mean are very close to the true values. Those obtained for the variance present larger errors. This is natural given that the variance calculation accumulates the errors from the calculations of the 1st and 2nd moments centered on the origin. In short, given the errors observed, it can be concluded that the results obtained through $B_C^D(t)$ are satisfactory.

It should also be noted that, in principle, the values obtained can be improved by decreasing $\Delta t$ (accuracy) and $\Delta p$ (precision). And we say in principle because the program running is very long and this slowness increases with the decrease in $\Delta t$ and $\Delta p$.

## 4 The $M|D|\infty$ Queue Busy Cycle Distribution Function Computation

A program similar to the one presented in the previous section can also be used to calculate the $M|D|\infty$ system busy cycle distribution function, since its Laplace transform is given in round form. In the previous program, simply use the expressions (1,9), (1.12) and (1.13) instead of the counterparts considered in it. We do not present it here to avoid making this text tedious.

---

[4]In fact, as the first customer has a service duration equal to $\alpha$ , the probability of the busy period lasting less than $\alpha$ is zero. The probability of being exactly $\alpha$ is the probability that the system will be empty when the first customer leaves the system, that is: $e^{-\rho}$.

The values of α, λ, Δ*t* and Δ*p* must be specified and also the values of *t* for which the values of $Z(t)$, called $Z^c(t)$, are wanted. The following calculations were performed:

- 1. $\alpha = 0$ and $\lambda = 1$ (Table 4.1)
- 2. $\alpha = 1$ and $\lambda = 1$ (Table 4.2)
- 3. $\alpha = 1$ and $\lambda = 2$ (Table 4.3)
- 4. $\alpha = 2$ and $\lambda = 1$ (Table 4.4)

The values of α, λ, Δ*t* and Δ*p* must be specified and also the values of *t* for which the values of $Z(t)$, called $Z^c(t)$, are wanted.

### Table 4.1

$\alpha = 0 \ \lambda = 1 \ \rho = 0 \ \Delta t = 0.01 \ \Delta p = .001$

| t | $Z^c(t)$ | Poisson Process |
| --- | --- | --- |
| 0 | 0.00020928263 | 0.000… |
| .5 | 0.39354845 | 0.39346934 |
| 1 | 0.63201874 | 0.632120559 |
| 1.5 | 0.77676630 | 0.77686984 |
| 2 | 0.86456292 | 0.864664717 |
| 2.5 | 0.91781115 | 0.917915001 |
| 3 | 0.95011103 | 0.95021212932 |
| 3.5 | 0.96969878 | 0.969802617 |

### Table 4.2

α=1 λ=1 $\rho = 1$ Δ t= 0.01 and Δ p=.001

| t | $Z^c(t)$ |
| --- | --- |
| .5 | 0.00070788896 |
| 1 | 0.00078194999 |
| 1.5 | 0.18467983 |
| 2 | 0.36851909 |
| 2.5 | 0.53561949 |
| 3 | 0.66881525 |
| 3.5 | 0.76919734 |
| 4 | 0.84198290 |
| 4.5 | 0.89332950 |
| 5 | 0.92884773 |
| 5.5 | 0.95303684 |
| 6 | 0.96932029 |
| 6.5 | 0.98016983 |
| 7 | 0.98734205 |
| 7.5 | 0.99205017 |
| $E[Z] = 2.718281829$ | $VAR[Z] = 1.9444392442$ |
| $E[Z^c] = 2.605018789$ | $VAR[Z^c] = 1.875647136$ |
| $ERROR = 4\%$ | $ERROR = 3.5\%$ |

**Table 4.3**.

α=1 λ=2 $\rho = 2$  Δ t= 0.01 and Δ p=.001

| t | $Z^c$(t) |
|---|---|
| 0.5 | 0.00038790601 |
| 1 | 0.00045109048 |
| 1.5 | 0.13572108 |
| 2 | 0.27099844 |
| 2.5 | 0.39718168 |
| 3 | 0.50513958 |
| 3.5 | 0.59509700 |
| 4 | 0.66922503 |
| 4.5 | 0.72997826 |
| 5 | 0.77964925 |
| 5.5 | 0.82022225 |
| 6 | 0.85335999 |
| 6.5 | 0.88039940 |
| 7 | 0.92047130 |
| 7.5 | 0.92047894 |
| 8 | 0.93518191 |
| 8.5 | 0.94718128 |
| 9 | 0.95697385 |
| 9.5 | 0.96496373 |
| 10 | 0.97148519 |
| 10.5 | 0.97680729 |

| 11 | 0.98115152 |
|---|---|
| 11.5 | 0.96469930 |
| 12 | 0.98759257 |
| 12.5 | 0.98995178 |
| 13 | 0.99188309 |
| 13.5 | 0.99344980 |
| 14 | 0.99473917 |
| $E[Z] = 3.69452805$<br>$E[Z^c] = 3.606224458$<br>$ERROR = 2.4\%$ | $VAR[Z] = 6.260481408$<br>$VAR[Z^c] = 5.358674148$<br>$ERROR = 14\%$ |

**Table 4.4**

α=2 λ=1 $\rho = 2$ Δ t= 0.01 Δ p=.001

| $t$ | $Z^c$(t) |
|---|---|
| 0.5 | 0.00039526703 |
| 1 | 0.00039531649 |
| 1.5 | 0.00039744257 |
| 2 | 0.00042999497 |
| 2.5 | 0.0068082088 |
| 3 | 0.13566480 |
| 3.5 | 0.20333376 |
| 4 | 0.27105104 |
| 4.5 | 0.33643096 |
| 5 | 0.39722785 |
| 5.5 | 0.45344632 |
| 6 | 0.50523263 |
| 6.5 | 0.55233818 |
| 7 | 0.59518069 |
| 7.5 | 0.63407224 |
| 8 | 0.66930794 |
| 8.5 | 0.70120662 |
| 9 | 0.73005634 |
| 9.5 | 0.75615197 |
| 10 | 0.77973318 |
| 10.5 | 0.80105113 |
| 11 | 0.82031202 |
| 11.5 | 0.83771467 |
| 12 | 0.85343867 |
| 12.5 | 0.86764937 |
| 13 | 0.88047999 |
| 13.5 | 0.89207541 |
| 14 | 0.90255320 |
| 14.5 | 0.91201680 |
| 15 | 0.92056465 |
| 15.5 | 0.92828899 |
| 16 | 0.93526571 |
| 16.5 | 0.94157290 |
| 17 | 0.94726365 |
| 17.5 | 0.95241045 |
| 18 | 0.95705801 |
| 18.5 | 0.96125179 |
| 19 | 0.96504825 |
| 19.5 | 0.96847575 |
| 20 | 0.97157025 |
| 20.5 | 0.97437018 |
| 21 | 0.97689431 |
| 21.5 | 0.97917509 |
| 22 | 0.98124003 |
| 22.5 | 0.98309797 |
| 23 | 0.98477888 |
| 23.5 | 0.98630297 |
| 24 | 0.98767584 |
| 24.5 | 0.98891764 |
| 25 | 0.99003869 |
| 25.5 | 0.99104917 |
| 26 | 0.99196279 |
| 26.5 | 0.99279278 |
| 27 | 0.99353820 |
| $E[Z] = 7.389056099$<br>$E[Z^c] = 7.200722486$<br>$ERROR = 2.5\%$ | $VAR[Z] = 25.04192563$<br>$VAR[Z^c] = 20.69584719$<br>$ERROR = 17\%$ |

As for the goodness of the obtained results, it is tested computing the errors of $E[Z^c]$ and $VAR[Z^c]$, computed after them, in relation with the true values of $E[Z]$ and $VAR[Z]$ that are available for this queue system. The exception is the first experience where, with α=0, the situation is the one of a pure Poisson express. So, the results obtained (2nd column in Table 4.1) are compared with the Poisson process ones (3rd column in Table 4.1). Generally, the $Z^c$ values fit well.

## 5 Conclusions

The results obtained attest to a reasonable performance of the algorithm. This performance is very dependent on the precision and accuracy chosen, and your careful choice can improve it. In other words, the program must be fine-tuned before running.